\documentclass[twocolumn,showpacs,preprintnumbers,amsmath,amssymb,prl,superscriptaddress]{revtex4}

\usepackage{graphicx}
\usepackage{dcolumn}
\usepackage{bm}

\begin{document}

\title{The self-assembly and evolution of homomeric protein complexes}
\author{Gabriel Villar}
\affiliation{Physical and Theoretical Chemistry Laboratory, 
  Department of Chemistry, University of Oxford, South Parks Road, 
  Oxford, OX1 3QZ, United Kingdom}
\author{Alex W.~Wilber}
\affiliation{Physical and Theoretical Chemistry Laboratory, 
  Department of Chemistry, University of Oxford, South Parks Road, 
  Oxford, OX1 3QZ, United Kingdom}
\author{Alex J.~Williamson}
\affiliation{Physical and Theoretical Chemistry Laboratory, 
  Department of Chemistry, University of Oxford, South Parks Road, 
  Oxford, OX1 3QZ, United Kingdom}
\author{Parvinder Thiara}
\affiliation{Physical and Theoretical Chemistry Laboratory, 
  Department of Chemistry, University of Oxford, South Parks Road, 
  Oxford, OX1 3QZ, United Kingdom}
\author{Jonathan P.~K.~Doye}
\thanks{Author for correspondence}
\affiliation{Physical and Theoretical Chemistry Laboratory, 
  Department of Chemistry, University of Oxford, South Parks Road, 
  Oxford, OX1 3QZ, United Kingdom}
\author{Ard A.~Louis}
\affiliation{Rudolf Peierls Centre for Theoretical Physics, 
 University of Oxford, 1 Keble Road, Oxford, OX1 3NP, United Kingdom}
\author{Mara N.~Jochum}
\affiliation{Physical and Theoretical Chemistry Laboratory, 
  Department of Chemistry, University of Oxford, South Parks Road, 
  Oxford, OX1 3QZ, United Kingdom}
\author{Anna C. F.~Lewis}
\affiliation{Physical and Theoretical Chemistry Laboratory, 
  Department of Chemistry, University of Oxford, South Parks Road, 
  Oxford, OX1 3QZ, United Kingdom}
\author{Emmanuel D.~Levy}
\affiliation{MRC Laboratory of Molecular Biology, Hills Road, Cambridge CB2 0QH,
United Kingdom}

\date{\today}

\begin{abstract}
We introduce a simple ``patchy particle'' model to study the thermodynamics
and dynamics of self-assembly of homomeric protein complexes. Our
calculations allow us to rationalize recent results for dihedral complexes.
Namely, why evolution of such complexes naturally takes the system into
a region of interaction space where 
(i) the evolutionarily newer interactions are weaker, 
(ii) subcomplexes involving the stronger interactions are observed to be 
thermodynamically stable on destabilization of the protein-protein 
interactions and 
(iii) the self-assembly dynamics are hierarchical with these same 
subcomplexes acting as kinetic intermediates.
\end{abstract}
\maketitle

\section{Introduction}

\begin{figure}
\includegraphics[width=8.4cm]{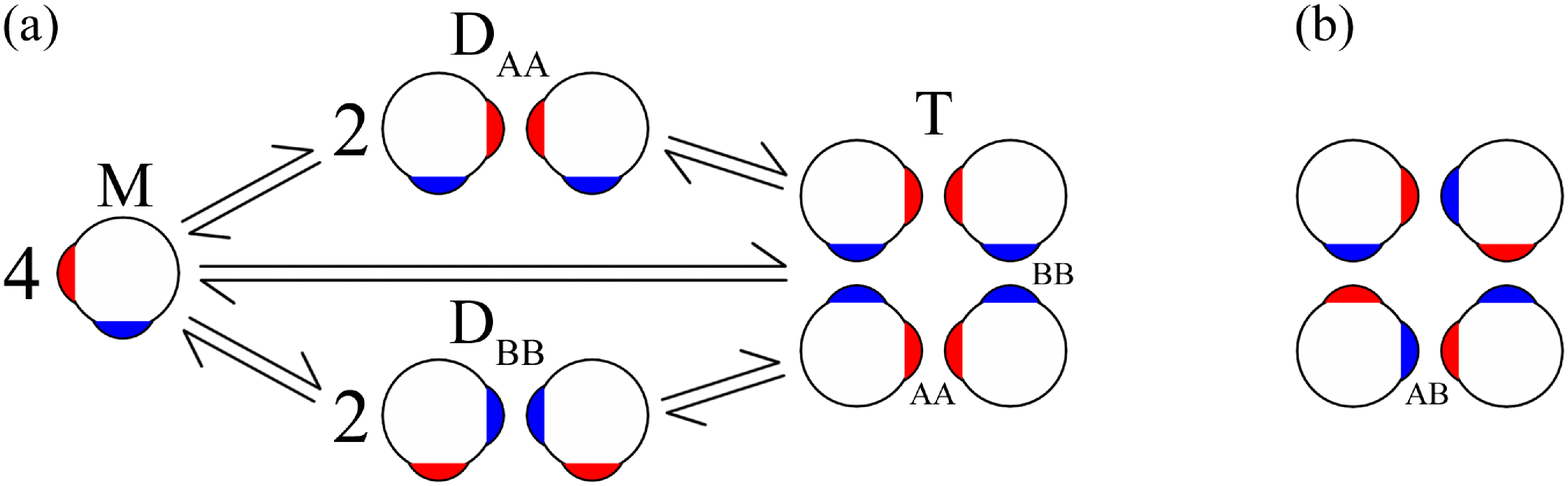}
\caption{\label{fig:structures}(Colour Online) 
Schmatic depiction of (a) a $D_2$ tetramer and the possible equilibria involved in its formation, and (b) a $C_4$ tetramer.
}
\vskip -0.3cm
\end{figure}

A large proportion of proteins are not monomeric {\it in vivo}, 
but instead 50-70\% of those of known structure
exist as homomeric protein complexes \cite{Levy06d,Goodsell00}. 
These complexes are usually symmetrical with each protein in an identical 
environment \cite{Goodsell00}. 
This latter constraint limits the number of possible symmetries
for such complexes. Thus, for homomeric tetramers, the two possible symmetries
that obey this rule are cyclic ($C_4$) or dihedral ($D_2$) 
(Fig.\ \ref{fig:structures}). The $C_4$ geometry involves only one type
of interaction, whereas the $D_2$ complex involves at least two 
self-complementary interactions. 
Interestingly, dihedral complexes are over ten times more abundant
than cyclic complexes with the same number of subunits \cite{Levy08}. 
The origin of this preference seems to be evolutionary, namely because
self-complementary interactions are easier to generate
\cite{Lukatsky06,Andre08} 
and because the evolution of 
dihedral complexes from a monomer does not have to proceed in a single step,
e.g.\ a $C_2$ dimer can be an intermediate on the evolutionary pathway to a 
$D_2$ tetramer. 

Insights into the evolution of homomeric protein complexes have
recently come from a study that compared the complexes adopted by
homologous proteins \cite{Levy08}. 
Although, for most homologues the quaternary structure is conserved, 
those cases where it is not conserved can tell us something about the
evolutionary relationships between complexes of different symmetry.
So, for example, it was found that where a tetramer
shared an evolutionary relationship with a dimer, the tetramer always had
$D_2$ symmetry and in the majority of cases 
the dimer interface was conserved in the tetramer, 
supporting the postulated role of the dimer as an evolutionary intermediate.
The study went on to show that the evolutionarily older interface 
was usually larger \cite{Levy08}, 
and so presumably had a stronger interaction strength.
In addition, mass spectrometry revealed that when the disassembly of 
dihedral complexes was induced by changing the solution conditions 
(e.g.\ through the addition of denaturant) stable subcomplexes involving 
the larger interfaces were detected in the majority of cases \cite{Levy08}.
Thus, the static structure of a dihedral complex, in particular the ratios
of the interface areas, can provide insight into the evolution and assembly
of the complex.

\begin{figure}
\includegraphics[width=8.4cm]{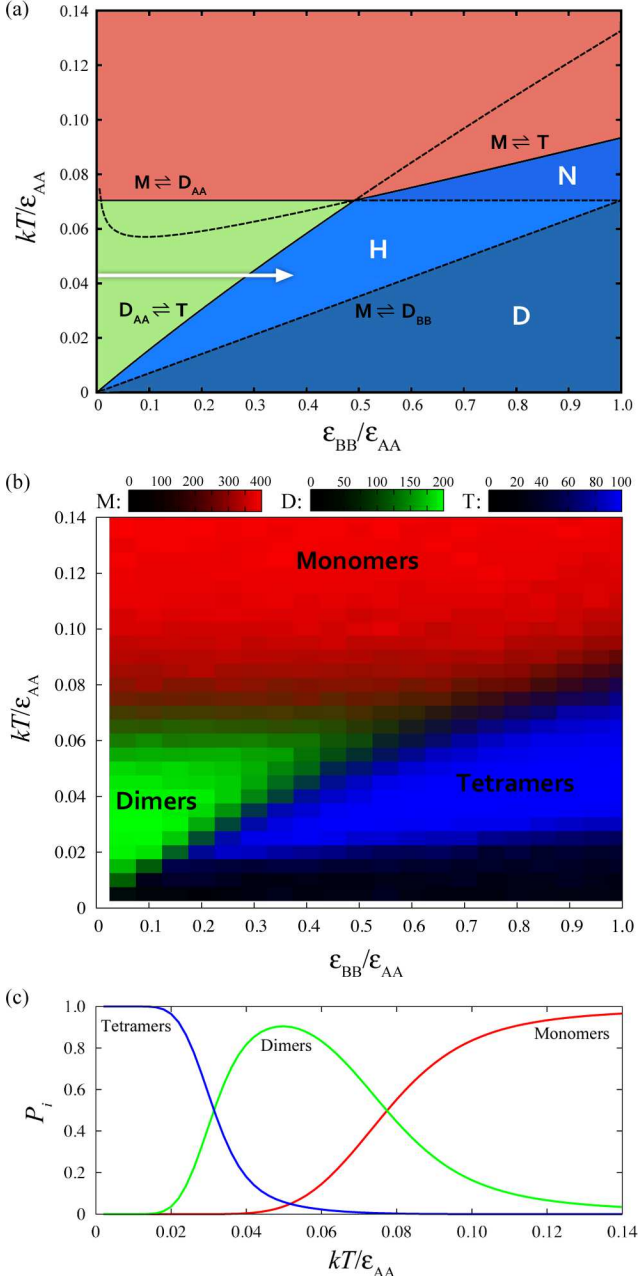}
\caption{\label{fig:FED}(Colour Online) 
(a) Free-energy diagram showing the dependence of the
most stable state of the $D_2$ tetramer system
on temperature and relative interaction strengths.
The lines 
indicate where the equilibrium constants for the
association reactions in Fig.\ \ref{fig:structures}(a) are one, and are solid
when the equilibrium is between the two most stable forms, and dashed otherwise.
The diagram is shaded according to whether monomers, dimers or 
tetramers are most stable.
The arrow indicates a possible evolutionary path from
a dimeric to a tetrameric complex.
(b) Dependence of the final yields of monomers, dimers and tetramers from 
our dynamics simulations on 
$kT/\epsilon_{AA}$ and $\epsilon_{BB}/\epsilon_{AA}$.
Each pixel represents the result of a separate simulation, where each
simulation started from a random configuration of the 400 particles 
and was $10^8$ steps long.
(c) Equilibrium probability of a particle being monomeric, or in a dimer
or tetramer at $\epsilon_{BB}/\epsilon_{AA}=0.2$.
}
\end{figure}

Here, we provide a framework for understanding these results
for dihedral complexes by studying the thermodynamics and dynamics of assembly
for a simple model of the complexes, particularly focussing on the role played 
by the relative strengths of the different interactions.

As the proteins in oligomeric complexes interact through highly specific contact
surfaces or ``patches'', we model the proteins using a patchy particle 
model used previously \cite{Wilber07,Doye07}, but with the introduction of an 
additional torsional component to the potential that ensures that
the patches must not only point at each other to interact strongly, 
but also have the correct relative orientation.  
In this model, the repulsion between the particles is based upon an isotropic 
Lennard-Jones potential:
\begin{equation}
V_{\rm LJ}(r) = 4\epsilon_{\rm ref}\left[ \left( \frac{\sigma_{LJ}}{r}
    \right)^{12} - \left( \frac{\sigma_{LJ}}{r} \right)^{6} \right], 
\label{eqn:LJ} 
\end{equation}
but where the attraction is modulated by an orientational term,
$V_{\rm ang}$.  Thus, the complete potential is
\begin{equation}
\label{eq:potential}
V({\mathbf r_{ij}},{\mathbf \Omega_i},{\mathbf \Omega_j})=\left\{
    \begin{array}{ll}
       V_{\rm LJ}(r_{ij}) & r_{ij}<\sigma_{\rm LJ} \\
       V_{\rm LJ}(r_{ij})
       V_{\rm ang}({\mathbf {\hat r}_{ij}},{\mathbf \Omega_i},{\mathbf \Omega_j})
                       & r_{ij}\ge \sigma_{\rm LJ}, \end{array} \right.
\end{equation}
where ${\mathbf \Omega_i}$ is the orientation of particle $i$, and
\begin{eqnarray}
V_{\rm ang}
({\mathbf {\hat r}_{ij}},{\mathbf \Omega_i},{\mathbf \Omega_j})
&=&\max\left[\frac{\epsilon_{\alpha\beta}}{\epsilon_{\rm ref}} 
\exp\left(-{\theta_{\alpha ij}^2\over 2\sigma_{\rm pw}^2}\right) \right. \nonumber\\
&&\left.
\exp\left(-{\theta_{\beta ji}^2\over 2\sigma_{\rm pw}^2}\right) 
\exp\left(-{\phi^2\over 2\sigma_{\rm tor}^2}\right)\right]
\end{eqnarray}
where 
$\theta_{\alpha ij}$ is the angle between the normal to patch $\alpha$ 
on particle $i$ and the interparticle vector $\mathbf r_{ij}$,
$\phi$ is a torsional angle, and the `$\max$' selects 
the pair of patches that have the strongest interaction for the current 
geometry. 
$\sigma_{\rm pw}^{-1}$ and $\sigma_{\rm tor}^{-1}$ are measures 
of the specificity of the patch-patch interactions, and 
which we here choose to be the same for all patches
\footnote{Throughout we use $\sigma_{\rm tor}/2=\sigma_{\rm pw}=0.475$
and a number density of $0.15\,\sigma_{LJ}^{-3}$.
These values were chosen because they lead to efficient assembly of 
tetramers at $\epsilon_{AA}=\epsilon_{BB}$.}. 
By contrast, we allow the well-depth of the patch-patch interactions, 
$\epsilon_{\alpha\beta}$ to vary 
($\epsilon_{\rm ref}=\max[\epsilon_{\alpha \beta}]$).
We also assume that interactions between pairs of patches not in contact in a
complex to be zero (e.g.\ $\epsilon_{AB}=0$ for $D_2$ tetramers).

To simulate the dynamics of our systems we use the `virtual move' Monte Carlo 
algorithm of Whitelam and Geissler \cite{Whitelam07} as this generates the 
diffusional behaviour expected of particles and clusters in solution.
To determine the thermodynamic properties of the system, we analytically 
calculate the partition functions for each state, 
as an ideal gas of clusters with rotational and vibrational degrees of 
freedom 
\footnote{G.\ Villar {\it et al.}, in preparation.}.
The vibrations are assumed to be harmonic and their frequencies
are calculated by diagonalization of the Hessian.

We first consider the case of $D_2$ tetramers, which for simplicity 
we choose to be planar. 
Fig.\ \ref{fig:FED}(a) shows the thermodynamics of the system
as a function of the ratios of the interaction strengths 
$\epsilon_{BB}/\epsilon_{AA}$. 
Note, due to the symmetry we need only consider the case 
where $\epsilon_{AA}>\epsilon_{BB}$.
At high temperature $T$ (or equivalently low $\epsilon_{AA}$) the system 
is monomeric. At $\epsilon_{BB}/\epsilon_{AA}=0$, as the system is cooled, 
it passes from a `gas' of monomers to a `gas' of dimers.
At non-zero $\epsilon_{BB}/\epsilon_{AA}$ a transition from
dimers to tetramers appears, and its transition temperature increases with
$\epsilon_{BB}/\epsilon_{AA}$ until a critical value of $\epsilon_{BB}/\epsilon_{AA}\approx 0.5$ is reached beyond which
dimers are no longer most stable for any value of $kT/\epsilon_{AA}$
\footnote{The position of this critical value of $\epsilon_{BB}/\epsilon_{AA}$ 
is insensitive to the potential parameters, being virtually 
constant as a function of density and patch specificity.}.
This critical value corresponds to the value of 
$\epsilon_{BB}/\epsilon_{AA}$ at the `triple point' where the 
three equilibrium lines in Fig.\ \ref{fig:FED}(a) meet.  

\begin{figure*}
\includegraphics[width=18cm]{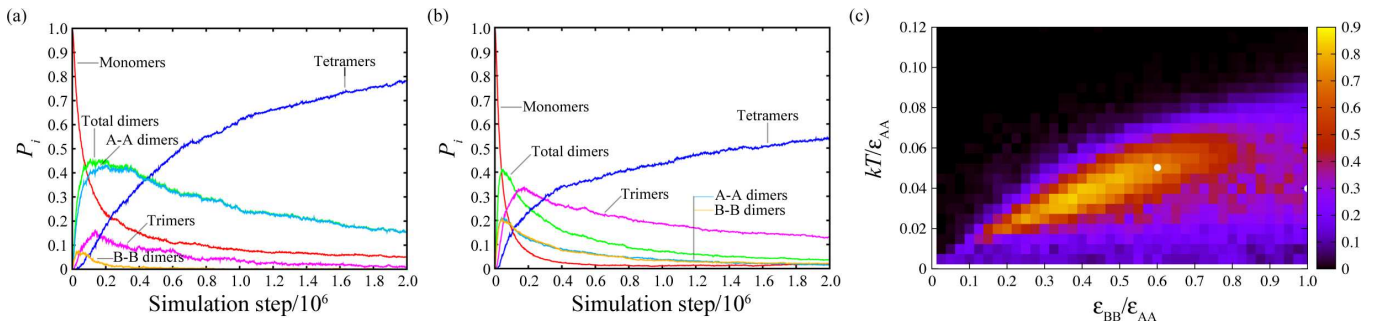}
\caption{\label{fig:dynamics}(Colour Online)
(a) and (b) Time dependence of the probability of a particle being in a
particular state. For (a) $\epsilon_{BB}/\epsilon_{AA}=0.6$ and 
$kT/\epsilon_{AA}=0.05$ and (b) $\epsilon_{BB}/\epsilon_{AA}=1$ and 
$kT/\epsilon_{AA}=0.04$. 
(c) Dependence of $h$, the hierarchicity index, on temperature and 
relative interaction strengths. The state points corresponding to (a) and (b) 
are marked by a white dot.
}
\end{figure*}

This disappearance of the dimeric state is easy to understand. In the monomer
to dimer transition the effective number of particles decreases by half, 
and the energy decreases by $\epsilon_{AA}$, whereas in the dimer to tetramer
transition, although the number of clusters again decreases by half, 
the energy decreases by $2\epsilon_{BB}$. 
Therefore, the  value of $\epsilon_{BB}/\epsilon_{AA}$
for which the monomer to $AA$ dimer and the $AA$ dimer to tetramer transitions 
occur at the same temperature is expected to be less than one, 
and is predicted to be 0.5, in agreement with Fig.\ \ref{fig:FED}(a), 
if the entropy change for both of these `dimerizations' is simply assumed to 
be equal.

The reliability of our analytical thermodynamic model is confirmed by 
comparisons to results obtained using the Wang-Landau 
algorithm \cite{Wang01b} and to 
our dynamics simulations. The kinetic yields of monomer, dimer and tetramer 
presented in Fig.\ \ref{fig:FED}(b) mirror the form of the free energy diagram 
(Fig.\ \ref{fig:FED}(a)), showing that near equilibrium is reached in the 
simulations except at very low temperatures where tetramers fail to form 
properly. In this latter case, the rate of cluster formation is so fast and
the rate of cluster breakup so slow that all the monomers are used up 
before complete tetramers can form. Instead, many particles get trapped in
trimers (e.g.\ Fig. \ref{fig:dynamics}(b)). 

We can use Fig.\ \ref{fig:FED}(a) to help us understand the evolution of a $D_2$
tetramer. 
The arrow in this figure shows a possible evolutionary pathways that takes
the system from a dimeric state (i.e.\ where $\epsilon_{BB}/\epsilon_{AA}=0$) 
to a region of interaction space where tetramers are stable.
The important thing to note is that the result of this evolution is a 
tetramer where the evolutionary newer $BB$ interactions are weaker. 
This conclusion is in agreement with Levy {\it et al.}'s observation 
that newer patches have smaller areas \cite{Levy08}, 
assuming that the interface area
is a reasonable proxy for estimating the interaction strength.

We can also use Fig.\ \ref{fig:FED}(a) to consider the effect of additives that
cause disassembly of protein complexes. If such additives destabilize all 
interactions equally, this process would correspond to a vertical pathway
in Fig.\ \ref{fig:FED}(a). Therefore, if a complex has a value of 
$\epsilon_{BB}/\epsilon_{AA}\lesssim 0.5$, 
a thermodynamically stable dimeric phase will be seen for some 
concentration of the denaturant. The equilibrium probabilities of being 
in the different states along one such pathway is illustrated in 
Fig.\ \ref{fig:FED}(c).
Again, as evolution will naturally take the system into a region to the left 
of the `triple point', our results are in agreement with the mass spectroscopic 
results of Ref.\ \onlinecite{Levy08} that detected subcomplexes involving the 
stronger interactions, and with more traditional studies of tetramer 
formation \cite{Jaenicke00,Powers03}. 
For example, Fig.\ \ref{fig:FED}(c) is very similar to results presented for
phosphofructokinase \cite{DevilleBonne89}.

Fig.\ \ref{fig:FED}(a) also allows us to think about the kinetic mechanisms
of tetramer formation. In particular, the region in which tetramers are
stable can be divided up into three subregions, depending on the stability 
of $AA$ and $BB$ dimers. In region $N$ neither dimers are stable with respect 
to monomers, and so all intermediates are unstable and there will be a
nucleation free energy barrier to tetramer formation.
In region $H$, $AA$ dimers, but not $BB$ dimers, are stable with respect to
monomers, and so $AA$ dimers will act as kinetic intermediates. In this region,
we therefore expect a hierarchical self-assembly mechanism to dominate, in 
which $AA$ dimers first form, and which in turn dimerize to form tetramers, 
rather than a mechanism which proceeds by sequential addition of monomers.
Finally, in region $D$ both $AA$ and $BB$ dimers are stable with respect to monomers
and so all pathways for tetramer formation are downhill in free energy.

To test these predictions, we first look at the time dependence of the 
probabilities of being in the different states. In region $D$,
the sequential passage from monomers to dimers
to trimers to tetramers is evident (Fig.\ \ref{fig:dynamics}(a)). 
However, the initial rapid rise in the 
number of tetramers slows down as monomers become depleted, and the further
formation of tetramers from trimers is dependent on cluster breakup 
(a relatively slow process) releasing additional monomers.
By contrast, in region $H$, $AA$ dimers are clear intermediates and there is a
steady growth of tetramers indicative of formation by dimer-dimer addition
(Fig.\ \ref{fig:dynamics}(b)).
Indeed, this plot has a very similar form to results for experimental studies 
on the rate of tetramer formation, 
such as for phosphoglycerate mutase \cite{Hermann83}
and lactate dehydrogenase \cite{Hermann83b}. 

\begin{figure}
\includegraphics[width=8.4cm]{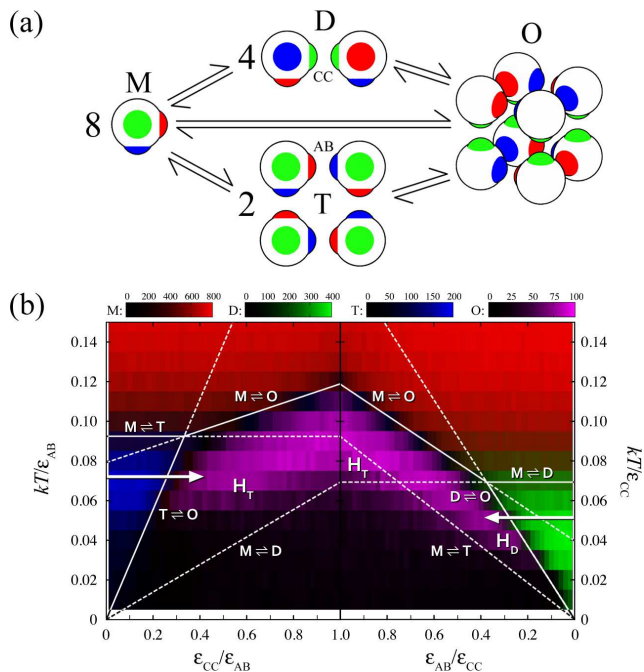}
\caption{\label{fig:octamer}(Colour Online) 
(a) Possible equilibria involved in the formation of a $D_4$ octamer.
(b) Free energy diagram showing the dependence of the most stable state
on temperature and interaction strengths, superimposed on the final yields
of monomers, dimers, tetramers and octamers in our dynamics simulations (800
particles, $10^8$ steps).
Note the change in the ordinate and abscissa in each half of the diagram.
}
\vskip -0.3cm
\end{figure}

We have further analysed how hierarchical the dynamics are by introducing
a `hierarchicity' index $h$ that we define as the fraction of tetramer forming 
events that occurred by dimer-dimer addition weighted by the fractional yield
of tetramers. It can be clearly seen that the region in Fig.\ 
\ref{fig:dynamics}(c) where $h$ is high corresponds well to region $H$ in 
Fig.\ \ref{fig:FED}(a).

The approach we have put forward in this paper for understanding the
formation of tetrameric complexes can be equally applied to complexes with
other numbers of subunits. To illustrate this, 
we show the free energy diagram for the formation of a $D_4$ octamer 
in Fig.\ \ref{fig:octamer}(b) calculated by a simpler version of the theory used for Fig.\ \ref{fig:FED}(a), superimposed on the yields of different
clusters obtained from our dynamics simulations. 
In our model for this system, the particles have
three patches, two types of interaction (AB and CC) 
and for simplicity the octamer has a cubic shape (Fig.\ \ref{fig:octamer}(a)).
There are two possible hierarchical pathways for octamer
assembly either via a $C_4$ tetrameric intermediate that is stabilized by the
AB interactions (region $H_T$) or via dimers stabilized by the CC interactions
(region $H_D$), and this expectation is confirmed by analyses of the dynamics 
in these regions.
In comparison to the results for the tetramer, it is noticeable that the
low temperature region in which incomplete assembly leads to poor yields 
extends to higher temperature; this is a general trend for the self-assembly of
more complex targets.
The arrows in Fig.\ \ref{fig:octamer}(b) illustrate potential evolutionary
pathways of an octamer from a cyclic tetramer or a dimer, and which will
again lead to complexes where the newer patches are weaker and where the
assembly is hierarchical. 

In this paper we have shown how an analysis of the dependence of the 
thermodynamics and kinetics of homomeric protein complexes on the relative 
interaction strengths can provide a framework for understanding how their 
properties are constrained by their evolution, in particular their asymmetry in
their interface areas and their hierarchical assembly.
Although we have focussed on tetrameric complexes, the approach is easily 
generalizable to larger complexes and leads to similar conclusions.
Thus, it would be very interesting if diagrams similar to the 
free energy diagrams of Figs.\ \ref{fig:FED}(a) and \ref{fig:octamer}(b) could 
be mapped out experimentally by studying in detail how the thermodynamics of
disassembly depends on the relative interface areas in a variety of protein
complexes.

\begin{acknowledgments}
The authors are grateful for financial support from the EPSRC and the Royal Society.
\end{acknowledgments}

\end{document}